\documentclass{vgtc}                          % final (conference style)
\ifpdf%                                % if we use pdflatex
  \pdfoutput=1\relax                   % create PDFs from pdfLaTeX
  \usepackage{graphicx}                % allow us to embed graphics files
  \DeclareGraphicsExtensions{.pdf,.png,.jpg,.jpeg} % for pdflatex we expect .pdf, .png, or .jpg files
\else%                                 % else we use pure latex
  \usepackage{graphicx}                % allow us to embed graphics files
  \DeclareGraphicsExtensions{.eps}     % for pure latex we expect eps files
\fi%

\graphicspath{{figures/}{pictures/}{images/}{./}} % where to search for the images

\usepackage{microtype}                 % use micro-typography (slightly more compact, better to read)
\PassOptionsToPackage{warn}{textcomp}  % to address font issues with \textrightarrow
\usepackage{textcomp}                  % use better special symbols
\usepackage{mathptmx}                  % use matching math font
\usepackage{times}                     % we use Times as the main font
\usepackage{cite}                      % needed to automatically sort the references
\usepackage{tabu}                      % only used for the table example
\usepackage{booktabs}                  % only used for the table example
\onlineid{0}

\vgtccategory{Research}

\vgtcinsertpkg

\title{Virtual Reality for medical education and training of Diabetic Foot}

\author{
Gabriel Riva\thanks{e-mail: gabriel.riva@unifesp.br }\\ %
\parbox{1.4in}{\scriptsize \centering ICT/UNIFESP, Universidade Federal de São Paulo, Brazil}%
\and Wellington Dores\thanks{e-mail: wellington.dores@unifesp.br}\\ %
     \parbox{1.4in}{\scriptsize ICT/UNIFESP, Universidade Federal de São Paulo, Brazil} %
\and Artur Damasio\thanks{e-mail: artur.damasio@unifesp.br }\\ %
     \parbox{1.4in}{\scriptsize ICT/UNIFESP, Universidade Federal de São Paulo, Brazil} %
\and Daniel Guimarães Cacione\thanks{e-mail: cacione@unifesp.br}\\ %
     \parbox{1.4in}{\scriptsize EPM/UNIFESP, Universidade Federal de São Paulo, Brazil} \\ %\\
\and \hspace{-1in}Joaquim Jorge\thanks{e-mail: jorgej@acm.org}\\ %
\parbox{3in}{\scriptsize INESC-ID, Instituto Superior Técnico, Portugal}
\and Ezequiel Zorzal\thanks{e-mail: ezorzal@unifesp.br}\\ %
\parbox{3in}{\scriptsize \centering INESC-ID, IST/UL, Portugal,  ICT/UNIFESP, Universidade Federal de São Paulo, Brazil}%
} %
\abstract{Diabetic Foot is one of the most common complications of Diabetes Mellitus and the leading non-traumatic cause of lower-limb amputations worldwide. These complications are preventable with early diagnosis and timely care. However, even in the context of expanding primary health care, this problem continues to increase, which suggests a gap in the training of primary health care professionals regarding the diagnosis and treatment of Diabetic Foot. This project proposes the development of a Virtual Reality simulator for training students and professionals in primary health care, aiming to collaborate in filling this gap. The application features gamification elements to increase user engagement. The context of medical care in primary care will be simulated with various clinical cases, including several virtual patients with different stages related to the Diabetic Foot, seeking to provide credible and distinct experiences to potential users. In addition, we aim to verify usability, effectiveness, and any side effects (cybersickness). Finally, we plan to conduct field studies with qualified students and professionals to identify the main benefits and obstacles to applying the technology.%
} % end of abstract

\CCScatlist{
  \CCScatTwelve{Computing methodologies}{Virtual reality}{}
}

\begin{document}


\firstsection{Introduction}

\maketitle

%% \section{Introduction} %for journal use above \firstsection{..} instead
\begin{figure*}[!htb]
 \centering % avoid the use of \begin{center}...\end{center} and use \centering instead (more compact)
 \includegraphics[width=\textwidth]{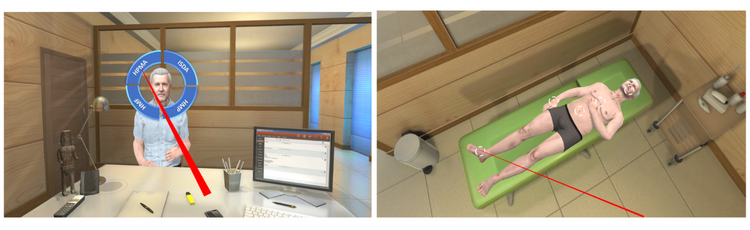}
 \caption{Use of Ray Casting techniques in interaction with patient care and physical examination (Source: adapted from www.mediactiv.com).}
 \label{fig:Interaction}
\end{figure*}
Diabetes Mellitus (DM) is a disease that affects about 7.9\% of the Brazilian population (approximately 16.8 million people) \cite{Binhardi2021} and is among the top five countries in the number of individuals with DM \cite{IDFDiabe77:online}. Still, there is evidence that about half of these individuals do not know they have the disease \cite{Sartorelli2003}.
The Diabetic Foot is among the most frequent complications of DM and its consequences can be traumatic for the individual's life and range from chronic wounds (present in 25\% of diabetics throughout their lives) and infections, to lower limb amputations \cite{Chastain2019}. Periodic examination of the feet provides early identification and timely treatment of the alterations found, thus enabling the prevention of a significant number of complications of the Diabetic Foot, such as ulcers and amputations \cite{Ahmad2016}. According to the Ministry of Health (Brazil), most non-traumatic lower limb amputations (about 60\%) are performed in patients with peripheral vascular disease and/or diabetes \cite{httpsbvs56:online}.

The training of general practitioners should include, among other topics and skills, the proper management of diseases that are common in the population and have the potential for serious complications and mortality. In the field of vascular surgery, the approach to the infected Diabetic Foot is highlighted due to the potential for complications, which may result in amputations that will impair the quality of life and work capacity of individuals, in addition to the possibility of mortality due to severe sepsis.

%It is noteworthy that possibly preventable occurrences are currently among the most frequent health complications caused by DM \cite{Silva2021}, even in a context in which, according to the Ministry of Health, DM patients rely on investments, receive medical assistance and care in the Unified Health System (SUS) \cite{Portalda97:online}.
%In this context, the qualification of primary care professionals regarding the early assessment and treatment of the Diabetic Foot is extremely relevant to contribute to the reduction of preventable cases of lower limb amputations and other complications arising from the lack of timely treatment.
%Measures related to conducting this patient profile are directly related to complications. Errors in recognizing the severity of the disease, use of inappropriate antibiotics in the initial approach to the patient, and delay in indicating an initial surgical approach, among other factors, are related to the increased incidence and level of lower limb amputations, length of inpatient care, stay in intensive care, and death.%length of hospital stay, stay in intensive care and death.

It is noteworthy that possibly preventable occurrences are currently among the most frequent health complications caused by DM \cite{Silva2021}, even in a context in which, according to the Ministry of Health, DM patients rely on investments, receive medical assistance and care in the Unified Health System (SUS) \cite{Portalda97:online}.
In this context, the qualification of primary care professionals regarding the early assessment and treatment of the Diabetic Foot is highly relevant to contribute to the reduction of preventable cases of lower limb amputations and other complications arising from the lack of timely treatment.
Measures related to conducting this patient profile are directly related to complications. For example, errors in recognizing %the severity of 
the disease severity, use of inappropriate antibiotics in the initial %approach to the patient,
treatment, and delay in %indicating an initial 
prescribing a surgical procedure, among other factors, are related to increased incidence of lower limb amputations, length of inpatient care, stay in intensive care and death.

Most of the time, the general practitioner (in the office, outpatient clinic, or emergency room) is one of the first professionals to approach the patient, formulating the diagnosis and conduct. In practice, if this professional on the “front line” manages to make an adequate diagnosis and appropriate conduct, the patient decreases the probability of complications. However, because this professional does not follow the evolution of the patients, he does not know the details of the complications and the direct consequences of the initial procedures.
Thus, the ideal teaching approach for general training practitioners on this topic should include a reasonable number of clinical scenarios, with patients with different disease severity, approaches, and outcomes. In addition, the evolution of these patients and the complications they present later are important in learning.

To contemplate these important aspects, the use of simulated approaches on the infected diabetic foot is highlighted. In traditional teaching, simulation is carried out through a discussion of cases at the bedside or presented by students and residents with discussion among graduates. This simulation, however, removes the possibility of individual decision-making and the direct consequences of the adopted conducts, in addition to being restricted to the number of cases available in the service.

Technology in general has been a great ally of medicine, especially when it comes to teaching and training. Recently, active teaching-learning methodologies based on interactive technologies, such as Virtual Reality (VR) and Augmented Reality (AR), have been used both to support the teaching and training of health professionals and to support them during surgical procedures \cite{Zorzal2020,Zorzal2021}.

Concomitantly, gamification has been used as an artifact to apply concepts of logic and game elements in the pedagogical process, making learning more attractive and challenging. According to Ogawa et al. (2015) \cite{NunesOgawa2016}, gamification is a process by which game elements are used in a non-game context. %other contexts.
Basten (2017) \cite{basten2017gamification} states that gamification can motivate people to learn in a different scenario from electronic games, popularly spread in recent decades. According to the author, gamified applications have been used for the most diverse learning problems. Gentry et al. (2019) \cite{gentry2019serious}, evaluated the effectiveness of using gamification for the education of health professionals compared to traditional learning. In this systematic review of the literature, the results of gamification were superior, showing evidence with a moderate to high level of certainty about its effectiveness concerning traditional learning. There was also evidence of small to large magnitude regarding the acquisition of improved skills compared to traditional learning. 

%The gamified approach using VR as a form of realistic simulation, allows the participant to be exposed to many scenarios, of various complexities, without being at the mercy of the availability of cases in the service, being able to glimpse and analyze the direct consequences of the conducts, without cost to the patient's life or even a medical structure (physical and available medical personnel) that can be accessed at any time and place are the main strengths of this approach. Currently, with restrictions on circulation due to the COVID-19 pandemic, approaches using technologies such as VR can be a valuable instrument in similar current and future situations \cite{Moreira2022}.

The gamified approach using VR as realistic simulation allows participants to be exposed to many scenarios, of various complexities, without being at the mercy of the availability of cases in the service, being able to glimpse and analyze the direct consequences of their conduct without endangering the patient's life or even medical infrastructure (physical and medical personnel) that can be accessed at any time and place are the main strengths of this approach. Furthermore, given restrictions on circulation due to the COVID-19 pandemic, approaches using VR technologies can be valuable in similar current and future situations \cite{Moreira2022}.

\begin{figure*}[!htb]
 \centering % avoid the use of \begin{center}...\end{center} and use \centering instead (more compact)
 \includegraphics[width=\textwidth]{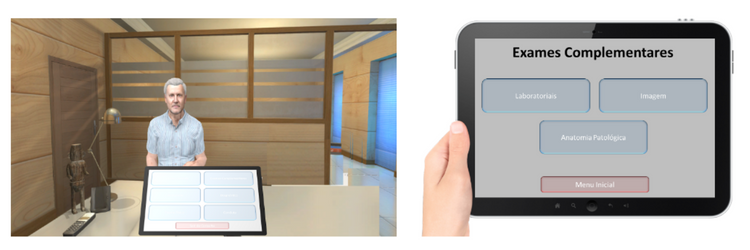}
 \caption{Interaction using virtual control through a tablet in a virtual environment (Source: adapted from www.mediactiv.com).}
 \label{fig:techniques}
\end{figure*}

In short, this project will have two main results: new scientific results, beyond the state of the art and established practice in traditional medical training; and a gamified simulator with VR producing proofs of concept that can be used and explored by users. The possibility of disseminating the produced results, in addition to the experience of the development of this technology, will be the main contribution of this project. %in the production of this technology will be the main contribution of this project.

%The team involved presents a line of research dedicated to teaching vascular surgery at EPM/UNIFESP and medical residency through the development of low-cost simulators, including funding from government agencies (FAPESP). In addition, medical teaching at EPM/UNIFESP is carried out in public hospitals (Hospital São Paulo/HU of UNIFESP and Hospital Vila Maria) with a marked presence of complications resulting from the Diabetic Foot and, therefore, with a volume of cases and proven expertise in the treatment of the infected diabetic foot. The team also has expertise in the development of VR applications in the medical field. The participation of both researchers from the Interactive Applications Lab - ICT/UNIFESP, of which the proponent is one of the coordinators, as well as the participation of the senior international researcher will be essential and will contribute to the effective development of the project.

There are not yet many studies in the literature relating the application of Virtual Reality technologies to DM in a healthcare context, which highlights the relevance of this work. Exploring the patient education side, Neira-Tovar et al\cite{tovar2017} present the development of a VR video game that aims to motivate exercise and combat depression, thus preventing some of the consequences of DM. Singleton et al\cite{Singleton2022} made use of a non-immersive VR simulation to teach nursing students about the treatment of diabetics patients and, in a randomized controlled trial, found that this method was significantly better than normative methods in terms of hypoglycemia knowledge. Collaborating with Oxford Medical Simulation, Mallik et al\cite{Mallik2021} developed fully immersive VR clinical scenarios related to DM and through Kirkpatrick's training evaluation model applied to thirty-nine non-specialist clinicians from 2 hospitals in the UK found an increased level of confidence in managing diabetes emergencies after the training sessions.

\section{Material and methods}

%\subsection{Simulator development}
The VR simulator will be developed based on the lifecycle model proposed by Preece et al. (2002) \cite{preece2002process}. This lifecycle was chosen because project requirements can be revised using inspection assessment results. Additionally, prototyping is a useful technique for facilitating user feedback on designs at all stages. To define the requirements, in addition to in-depth studies through the RSL carried out in the previous stage, we will consider interviews with the target audience, to deepen the understanding of how real experiences of care in primary care take place and the forms of interaction between the health professional and the patient, collecting information about all stages of the care process. This understanding is essential for defining the requirements for creating an immersive application that is as close as possible to the real experience.

%Application development in the prototyping, minimum viable product (MVP) and final product phases will be carried out using the Unity Engine tool. The 3D modeling of the scenarios, objects, and avatars that will compose the application will be developed using the Blender 3D tool.
%Once the simulator requirements are defined, the next step will be the development of prototypes with different forms of interaction, aiming to find the best alternatives through the feedback of expert users. Feedback will be collected through tests and semi-structured interviews.
%As alternatives for interactions in the virtual environment, the following forms will be developed and tested: interaction with a virtual control, from a virtual tablet represented in the three-dimensional environment; selection and manipulation by ray casting and; the use of hand tracking or Touch controls (VR controller).

Application development in the prototyping, minimum viable product (MVP), and final product phases will be carried out using the Unity Engine tool. The 3D modeling of the scenarios, objects, and avatars that compose the application will be developed using Blender 3D.
Once the simulator requirements are defined, %the next step will be the 
we will develop prototypes with different interactions, aiming to find the best alternatives through expert user feedback, collected through tests and semi-structured interviews.
As alternatives for interactions in the virtual environment, the following forms will be developed and tested: interaction with a virtual tablet represented in the three-dimensional environment; selection and manipulation by ray casting; using hand tracking or VR controllers.

In the interaction with a virtual control (Figure \ref{fig:Interaction}), a three-dimensional model will be developed representing a tablet to be inserted in the simulator. Thus, intuitively, it can be manipulated by the user, presenting different types of information and control options. We will use the user hand tracking to access and interact with the virtual controller. Hand tracking will work using the inside-out cameras on the Oculus Quest 2 headset or similar. In this way, the headset will detect the position and orientation of the hands, in addition to the configuration of the user's fingers. This tracking technique is widely used in situations that need to leave the user's hands-free, as well as it is already a feature available on the Oculus Quest 2 headset and similar devices. With movements and gestures of the user's hands in the headset's field of view, it will be possible to select and perform specific interactions with the virtual control.

%The Ray Casting technique is a very common interaction in VR~\cite{Tu2019,Lu2020}. %The concept basically consists of 
%It entails aiming at a virtual object through a beam controlled by the user. As the ray hits the object, it can be selected and later manipulated, allowing interaction with objects that are out of the direct reach of the user. Figure \ref{fig:techniques} presents a sketch that we intend to develop as an alternative for interaction in the simulator. By pointing to a specific location, properly signaled, a virtual menu with the available options will be presented for the user to interact and make choices.
The Ray Casting technique is a pervasive interaction in VR~\cite{Tu2019,Lu2020}. It entails aiming at a virtual object through a beam controlled by the user. As the ray hits the object, it can be selected and later manipulated, allowing interaction with objects that are out of the direct reach of the user. Figure \ref{fig:techniques} presents a sketch we intend to develop as an alternative for interaction in the simulator. A virtual menu with the available options appears when pointing to a specific location, correctly signaled, so the user can interact and make choices.

In the interaction with the hands or by Touch control, the user will be able to use his own hands to select and manipulate the objects that are close in the virtual environment. In the same way that the interaction with the virtual controller will be implemented, we will use the available resources of the headset to track the user's hands. We intend to perform gestures such as pointing to select a specific object, as well as performing pinch-like movements with the fingers, to hold and move the objects available in the virtual environment. It should be noted that we intend to combine these interaction features with the Touch controls available on the Oculus Quest 2 headset or similar, in such a way that the headset automatically switches between using the hands or physical controls, if the user drops the controls or picks them up again. %Figure 4 presents an example of interaction using hands, from an Oculus Quest controller to represent a stethoscope.

%\begin{figure}[tb]
 %\centering % avoid the use of \begin{center}...\end{center} and use \centering instead (more compact)
 %\includegraphics[width=\columnwidth]{paper-count-w-2015-new}
 %\caption{Example of Interaction using the hands, through a Touch Oculus Quest control to represent a stethoscope (Source: VR Patients – www.vrpatients.com)}
 %\label{fig:sample}
%\end{figure}

%Figure 4 – Example of Interaction using the hands, through a Touch Oculus Quest control to represent a stethoscope (Source: VR Patients – www.vrpatients.com).

According to Edwards et al. (2016) \cite{edwards2016gamification}, the gamification techniques commonly applied are individual scores, ranking, rewards, avatars, badges, competitions, and challenges. Both researchers state that these techniques can be applied to several gamified systems, to motivate users to fulfill their tasks. Based on this statement, we intend to develop and incorporate the following gamification techniques into the simulator:

\begin{itemize}
\vspace{-1mm}
\item Individual scoring: Individual points will be awarded when users successfully perform activities. It should be noted that the score may be different for each type of activity;
\vspace{-2mm}
\item Loss aversion: The administrator can choose to use the loss aversion system and define the number of points that people can lose when they fail to reach specific simulation objectives;
\vspace{-2mm}
\item Rewards: The administrator will be able to configure the attribution of rewards according to some established rules. You can reward users via badges or other prizes when the user reaches a specific score or when they manage to finish a challenge;
\vspace{-2mm}
\item Time restriction: The simulator can be configured to restrict the time for each challenge.
\vspace{-2mm}
\item Ranking: The simulator will be able to present a classification table (ranking) with the individual score of each user.
\vspace{-2mm}
\end{itemize}

We intend to store user data, such as score, level, and preference, on a MySQL-enabled Web server to facilitate the retrieval of information for each user.

\subsection{Simulation steps}
%The simulation will take place in a virtual environment simulating a doctor's office with two main environments. The first environment will have a desk, chairs, and some objects on top of the desk. The second environment is where the physical exam will be simulated and will have a stretcher where the virtual patient can lie down and sit, in addition to the objects and instruments necessary for asepsis and carrying out the exams.
The simulation will occur in a virtual environment simulating a doctor's office with two primary settings. The first will have a desk, chairs, and some objects on top of the desk. The second is where the physical exam will be simulated and a stretcher where the virtual patient can lie down and sit, in addition to the objects and instruments necessary for asepsis and the exams.
%When starting the simulation, the user will be positioned on one side of the desk, and in front of him, sitting on the other side of the desk will be the avatar of the virtual patient of the chosen case. On the desk will be the tablet or virtual controller, which will be one of the forms of interaction to be tested. On this tablet, the user will be able to check the patient's file with information such as initials, age, sex, weight, height, and vital signs, in addition to the complaint and duration of the reported problem. This form will also be available on a clipboard on the table as an alternative to viewing written information. It will be possible to proceed to the next stages of the simulation, beginning with anamnesis (Figure \ref{fig:Anamnesis}), by pressing a button on the tablet or interacting with Ray Casting by pointing to the virtual patient. 

\begin{figure}[!b]
 \centering % avoid the use of \begin{center}...\end{center} and use \centering instead (more compact)
 \includegraphics[width=0.5\textwidth]{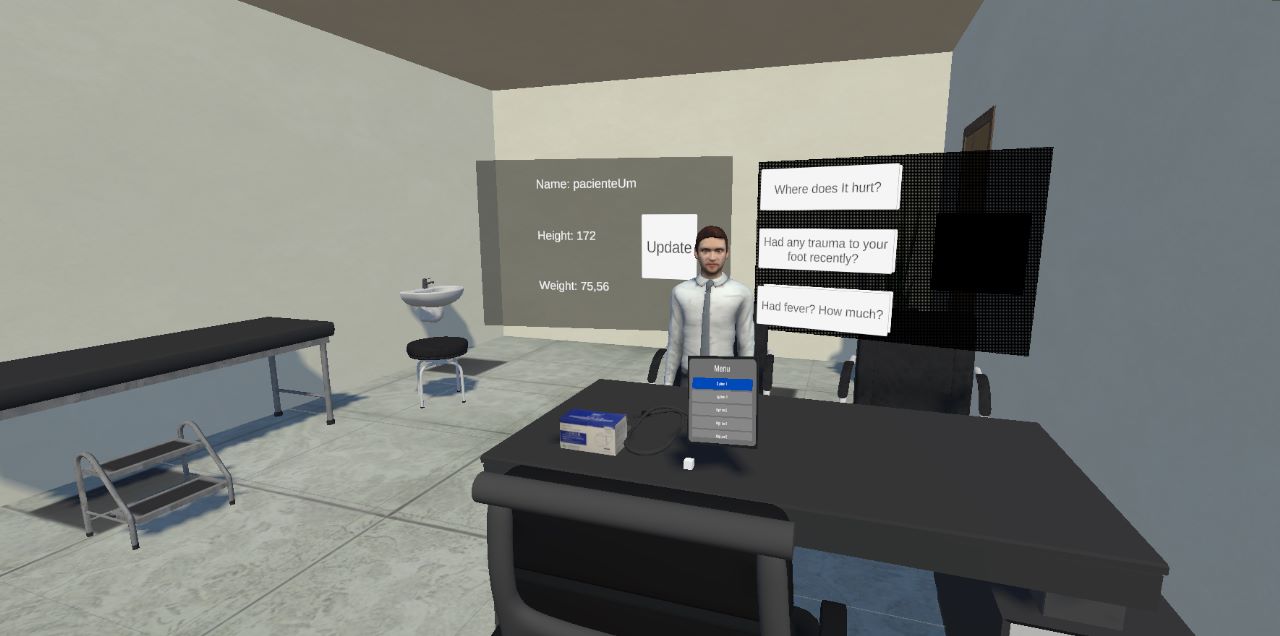}
 \caption{Anamnesis: This stage is the "interview" with the patient, where the doctor asks the patient questions to better understand their complaint and help in the diagnosis.}
 \label{fig:Anamnesis}
\end{figure}

When starting the simulation, the user will be positioned on one side of the desk, and in front of him, sitting on the other side of the desk, will be the avatar of the virtual patient of the chosen case. On the desk will be the tablet or virtual controller, which will be one of the forms of interaction to be tested. On this tablet, the user can check the patient's file with information such as initials, age, sex, weight, height, vital signs, and the complaint and duration of the reported problem. This form will also be available on a clipboard on the table as an alternative to viewing written information. It will be possible to proceed to the following simulation stages, beginning with anamnesis (Figure \ref{fig:Anamnesis}), by pressing a button on the tablet or interacting with Ray Casting by pointing to the virtual patient. 

%The simulation will consist of five sequential steps: Anamnesis, Physical Examination, Complementary Exams, Diagnosis, and Conduct and Unfolding(follow-up). Each step will be simulated through different forms of interaction and will generate a score for the gamification according to the user's performance. In the final step, a report of the user's performance will be displayed and the simulation will return to the Cases Menu screen.

The simulation will encompass five consecutive stages: Anamnesis, Physical Examination, Complementary Examinations, Diagnosis, and Follow-up. Each stage will be simulated through various forms of interaction and a score will be generated for gamification, reflecting the user's performance. The final stage will culminate in the display of a report on the user's performance, after which the simulation will return to the main Cases Menu screen.

\subsection{Evaluation of results}

To assess how users will perceive the features implemented in the simulator, in terms of usefulness and usability, we will use a think-aloud protocol to conduct qualitative evaluation sessions with users. In addition, a system usability scale (SUS) questionnaire \cite{brooke1996sus} will be applied to measure usability, and a NASA Task Load Index (NASA-TLX) \cite{hart1988development} questionnaire for perceived workload rates to assess prototype effectiveness. The effect of the VR headset will be evaluated during testing using quantitative metrics such as task completion, error rates, and time. The performance evaluation will analyze the learning results, the subjective evaluation of the post-training students, and the acquired proficiency in comparison with the current techniques. The ANOVA test \cite{Kaufmann2014} can be performed to detect statistically significant differences between the averages of the applications if the assumptions of homogeneity of variances and normal distribution of residuals were verified. Otherwise, the data will be analyzed using a non-parametric technique such as the Friedman test. After performing the appropriate test, a post hoc test should be applied to perform multiple pairwise comparisons (eg Tukey's or Dunn's tests). More details about the tests can be found in Bailey (2008) \cite{bailey2008design}. All statistical analyzes will be performed using the statistical software R \cite{rizzo2019statistical}.

\section{Conclusion}

%This work aims to develop a VR simulator to train healthcare professionals regarding the diagnosis and treatment of Diabetic Foot. The simulator will be developed using the Unity game engine optimized for the Oculus Quest platform, and gamification features elements. The user will interact with the simulator using virtual controls, ray casting, hand tracking and Touch controls. These immersive technologies can be a promising ally in the improvement of medical practices. And the results of this project allow the evaluation of the use of these technologies in a relevant medical problem: Diabetic Foot Infection. Also, this work opens up the development foundations for future applications of this kind in many other areas of medicine, with relevant impact on the teaching and care of patients over the medium to long term.    

This work aims to develop a VR simulator to train healthcare professionals regarding the diagnosis and treatment of Diabetic Foot. The simulator will be developed using the Unity game engine optimized for the Oculus Quest platform and gamification features elements. The user will interact with the simulator using virtual controls, ray casting, hand tracking, and Touch controls. These immersive technologies can be a good ally in improving medical practices. Moreover, the results of this project allow the evaluation of the use of these technologies in a relevant medical problem: Diabetic Foot Infection. Also, this work opens up the development foundations for future applications of this kind in many other areas of medicine, with relevant impact on the teaching and care of patients over the medium to long term.  

\acknowledgments{
This work is financed by National Research Council (CNPq), grant 423521/2021-7 and Portuguese Science Foundation under Grant 2022.09212.PTDC and supported by UNESCO Chair on AI\&XR.
}

\bibliographystyle{abbrv}

\bibliography{template}
\nocite{Pereira04,goncalves03,jorge2011sketch,correia05}
\end{document}